\documentclass[amsmath,amssymb,nofootinbib,11pt]{revtex4}
\pdfoutput=1
\usepackage{graphicx}
\usepackage{epsfig}
\usepackage{rotate}
\usepackage{amsmath}
\usepackage{amssymb}
\usepackage{amsfonts}
\usepackage{enumerate}
\usepackage{afterpage}
\usepackage{color}
\usepackage{amscd}
\usepackage{hyperref}

\def\be{\begin{equation}}
\def\ee{\end{equation}}
\def\bea{\begin{eqnarray}}
\def\eea{\end{eqnarray}}

\newcommand{\Nb}{\nabla}

\newcommand{\no}{\noindent}

\newcommand{\de}{\partial}

\renewcommand\a{\alpha}

\renewcommand\b{\beta}

\newcommand\m{\mu}
\newcommand\n{\nu}
\newcommand\g{\gamma}

\newcommand\s{\sigma}

\newcommand\Ost{Ostrogradsky }

\begin{document}

\title{Horndeski: beyond, or not beyond?}

\author{Marco Crisostomi$^a$,  Matthew Hull$^a$, Kazuya Koyama$^a$, and Gianmassimo Tasinato$^{b}$}

\affiliation{
	$^a$Institute of Cosmology and Gravitation,~University of Portsmouth, Portsmouth, PO1 3FX, UK\\
	$^b$Department of Physics, Swansea University, Swansea, SA2 8PP, UK
}

\hskip0.5cm
\begin{abstract}
\no Determining the most general, consistent scalar tensor theory of gravity is important for building models of inflation and dark energy. In this work we investigate the number of degrees of freedom present in the theory of beyond Horndeski.  We discuss how to construct the theory from the extrinsic curvature of the constant scalar field hypersurface, and find a simple expression for the action which guarantees the existence of the primary constraint
necessary to avoid the \Ost instability.
Our analysis is completely gauge-invariant.
However we confirm that, mixing together beyond Horndeski with a different order of Horndeski,
obstructs the construction of this primary constraint.
Instead, when the mixing is between actions of the same order, the theory can be mapped to Horndeski through a generalised disformal transformation. This mapping however is impossible with beyond Horndeski alone, since we find that the theory is invariant under such a transformation.
The picture that emerges is that beyond Horndeski is a healthy but isolated theory: combined with Horndeski, it either becomes Horndeski, or likely propagates a ghost.

\end{abstract}

\maketitle
\section{Introduction}

\no The question concerning the form of the most general, consistent scalar tensor theory of gravity is frequently being reconsidered due to its foremost physical importance for building models of inflation and dark energy \cite{Koyama:2015vza}. The problem is deeper than expected, and its theoretical investigation leads to intriguing surprises. Subtleties arise when considering derivative couplings of gravity to the scalar field, due to the non-linearity of the equations of motion, and the non trivial structure of the constraint equations. A systematic study of scalar-tensor theories as a modification of General Relativity (GR) started with the work of Brans and Dicke \cite{Brans:1961sx}, and subsequent developments mainly considered scalar-tensor Lagrangians with non-minimal couplings of the scalar to curvature, but containing at most first derivatives of the scalar field (see \cite{Fujii:2003pa} for a comprehensive review). 

The modern approach to scalar tensor theories started with the introduction of Galileons, the most general scalar field theory in flat spacetime with second order equations of motion \cite{Nicolis:2008in}. Galileon actions contain second order derivatives acting on the scalar field. They have very attractive phenomenological and theoretical features. In particular, they admit self-accelerating solutions that can explain the phenomenon of dark energy, and at the same time exhibit screening mechanisms, which ensure consistency with the tight bounds on small scale deviations from GR. 
Once Galileons are covariantised by naively replacing partial derivatives by covariant derivatives, the theory leads to third order derivatives in the equations of motion. A way to avoid them is to introduce counter terms, in the form of non-minimal couplings to gravity \cite{Deffayet:2009wt}. It was then shown that these theories can be further extended by including free functions of the scalar field and its kinetic term \cite{Deffayet:2009mn}. Remarkably, it was found that this theory was already proposed in 1974 by Horndeski \cite{Charmousis:2011bf,Kobayashi:2011nu} and this theory is now called the Horndeski theory \cite{Horndeski:1974wa}. 

The basic strategy to construct the Horndeski theory is to ensure that the equations of motion are of at most second order in derivatives. If the equations of motion contain higher order time derivatives, this generally leads to a ghost instability due to the Ostrogradsky theorem\footnote{See Ref.~\cite{Woodard:2006nt} for a review.}. However, it was suggested that -- despite the appearance of higher order derivatives --  it is still possible to avoid the Ostrogradsky ghost if the theory contains a hidden constraint and the equations of motion can be recast into a second order form, for example by field redefinitions. This possibility was studied in detail by Ref.~\cite{Zumalacarregui:2013pma, Gleyzes:2014dya, Gleyzes:2014qga,Gao:2014soa}. In Ref.~\cite{Gleyzes:2014dya, Gleyzes:2014qga}, the authors constructed theories in the so-called unitary gauge where the scalar field depends only on time, thus allowing it to be used as a clock. It was then found that there are two more terms with two arbitrary functions of the scalar field and its kinetic term that can be added to the Horndeski theory. Remarkably, these two terms can be obtained by naively covariantising the flat space Galileon in the form written with the anti-symmetric {Levi-Civita tensor}. Although the equations of motion contain third order time derivatives, it was argued that they do not lead to the Ostrogradsky instability, since constraints remove this additional degree of freedom. These terms are dubbed `beyond Horndeski'. 

There still remains the question of whether beyond Horndeski can be obtained from Horndeski by a suitable transformation of the metric. It is indeed possible to perform  generalised conformal and disformal transformations \cite{Bekenstein:1992pj}, that depend on the scalar field and its kinetic term, on Horndeski and obtain a theory which contains higher order derivatives \cite{Zumalacarregui:2013pma, Domenech:2015tca}. As long as the transformation is invertible, it does not change the physics. Thus, despite the appearance of higher order derivatives, the resultant theory propagates the same number of degrees of freedom as Horndeski. However, in this case, the theory can be mapped into Horndeski, so unless we introduce a coupling to matter, the new theory is nothing but Horndeski. Indeed, it has been shown that a beyond Horndeski term can be generated by a generalised disformal transformation from Horndeski \cite{Gleyzes:2014qga}. 

This brings us to the question of whether beyond Horndeski is really beyond Horndeski or not.  In this paper, we re-examine the problem of determining the number of degrees of freedom in the theory of beyond Horndeski \cite{Deffayet:2015qwa, Langlois:2015cwa, Langlois:2015skt}, without selecting any gauge.  We reformulate the theory using geometrical quantities defined with respect to a constant scalar field hypersurface. This representation gives a simple geometrical understanding of why the higher derivatives in beyond Horndeski do not lead to any additional degrees of freedom. We show this fact by demonstrating the existence of a primary constraint in the Hamiltonian formulation. We further discuss cases in which, when the theory is combined with Horndeski, this constraint is spoiled.
We finally study the conditions for mapping beyond Horndeski to Horndeski  by a generalised disformal transformation. 

This paper is organised as follows. In section \ref{sec-rev}, we introduce Horndeski and beyond Horndeski theories.
We formulate beyond Horndeski  using geometrical quantities defined with respect to the constant scalar field hypersurface.
In section \ref{sec-kin}, we derive kinetic terms for the metric and a scalar field focusing on its higher derivative terms. In section \ref{sec-prim}, we examine the existence of a primary constraint, which is necessary for the non-propagation of the additional mode due to the higher derivatives. In section \ref{sec-disf}, we study the possibility of mapping beyond Horndeski to Horndeski by a generalised disformal transformation. Section \ref{sec-disc} is devoted to conclusions. 

\section{The theories of Horndeski and beyond Horndeski}\label{sec-rev}

\no In this section we review the theories of Horndeski and beyond Horndeski, and reformulate the latter using geometrical quantities.

\subsection{Horndeski theory}

\no The scalar tensor theory of Horndeski is given by the action
\begin{eqnarray}
{}^{H} S=\int d^4 x\,\sqrt{-g}\,\left({}^{H} {\cal L}_2
+{}^{H} {\cal L}_3+{}^{H} {\cal L}_4+{}^{H} {\cal L}_5\right),
\end{eqnarray}
where the Lagrangian densities are
\begin{eqnarray}
{}^{H}{\cal L}_2&=&K(\phi, X) \,, \\
{}^{H}{\cal L}_3&=&G_3(\phi, X)\Box\phi \,,\\
{}^{H}{\cal L}_4&=&G_4(\phi, X)R
- 2 G_{4X}(\phi, X)\left[(\Box\phi)^2-\phi_{\mu \nu} \phi^{\mu \nu}\right] \,,
\\
{}^{H}{\cal L}_5&=&
G_5(\phi, X)G^{\mu\nu}\phi_{\mu \nu}
+\frac{1}{3}G_{5X}(\phi, X)\left[(\Box\phi)^3
-3\, \Box \phi\, \phi_{\mu \nu} \phi^{\mu \nu} 
+ 2\, \phi_{\mu \nu}\phi^{\nu \rho} \phi^{\mu}_{\rho}\right]\,.
\end{eqnarray}
$\phi_{\mu \nu} \equiv \nabla_{\mu} \nabla_{\nu} \phi$\, and $K$, $G_{3,\,4,\,5}$ are arbitrary functions of 
$\phi$ and $X$, defined as
\be
X\,\equiv\,\partial_\mu \phi\,\partial^\mu \phi\,.
\ee
This is the most general  scalar tensor theory of gravity, involving a single scalar field, that leads to covariant second order equations of motion. 

The cubic Horndeski theory is described by the Lagrangian ${\cal  L}_3$. In this case gravity becomes dynamical only through a mixing with the scalar field, a phenomenon dubbed kinetic gravity braiding, see \cite{Deffayet:2010qz}. In the present work, we focus on the quartic  ${\cal  L}_4$, and the quintic ${\cal L}_5$ Horndeski theories,  where the tensor spin-2 degrees of freedom  have  their  own kinetic terms. General Relativity is recovered by setting $K=G_3=G_5=0$, and $G_4\,=\,M_{\text{Pl}}^2/2$. 

In order to study the dynamics of theories with second order derivatives in the action, it is convenient to adopt the field redefinition introduced in Ref.~\cite{Langlois:2015cwa}. This is the natural covariant extension of the one used by \Ost in his pioneering work (see \cite{Woodard:2006nt}).
%
We define therefore a four vector
\begin{equation}
A_{\mu} \equiv \nabla_{\mu} \phi.
\end{equation}
Using $A_{\mu}$, the quartic and quintic Horndeski Lagrangians are rewritten as 
\begin{eqnarray}
{}^{H}{\cal L}_4&=&G_4(\phi, X)R
-2G_{4X}(\phi, X)\,
\delta^{\alpha}_{[\mu} \delta^{\beta}_{\nu]} \, \nabla_{\alpha} A^{\mu}
\nabla_{\beta} A^{\nu} \,,
\\
{}^{H}{\cal L}_5&=&
G_5(\phi, X)G^{\a}_{\mu} \, \nabla_{\a} A^{\mu}
+\frac{1}{3}G_{5X}(\phi, X)\,
\delta^{\alpha}_{[\mu} \delta^{\beta}_{\nu}  \delta^{\gamma}_{\rho]}\, \nabla_{\alpha} A^{\mu} \nabla_{\beta} A^{\nu}\nabla_{\gamma} A^{\rho} \,,
\end{eqnarray}
where  $[...]$ denotes anti-symmetrisation\footnote{Notice that in our definition there is no factor $1/n!$ in front of (anti-)symmetrised tensor.} and
\be
X = A_{\mu} A^{\mu} = A^2 \,.
\ee

\subsection{Beyond Horndeski theory}
\label{secBHT}

\no As stated above, the Horndeski system is the most general scalar tensor theory leading to covariant second order equations of motion for the scalar and tensor field, hence it certainly propagates at most three degrees of freedom. Intriguingly, it is conceivable that more general scalar tensor theories exist, whose equations of motion are higher order, but at the same time are characterised by constraints that remove additional, undesired degrees of freedom. A proposal in this direction is the theory of ``beyond Horndeski'', described by the following action
\begin{eqnarray}
{}^{BH} S=\int d^4 x\,\sqrt{-g}\,\left({}^{BH} {\cal L}_4+{}^{BH} {\cal L}_5\right),
\end{eqnarray}
where
\begin{eqnarray}
{}^{BH}{\cal L}_{4} &=&  F_4(\phi,\, X) \Big[
X \left( (\Box\phi)^2-\phi_{\mu \nu} \phi^{\mu \nu} \right)
-2 \left( \Box \phi\, \phi_{\mu}
\phi^{\mu\nu}\phi_{\nu} 
 - \phi_{\mu}\phi^{\mu\nu}
\phi_{\nu\rho}\phi^{\rho}\right) \Big], \\
{}^{BH}{\cal L}_{5} &=& 
F_5(\phi, \,X) 
\Big[
X \left( (\Box\phi)^3- 3\, \Box \phi\, \phi_{\mu \nu} \phi^{\mu \nu} 
+ 2 \phi_{\mu \nu}\phi^{\nu \rho} \phi^{\mu}_{\rho}
\right) \nonumber\\
&-& 3 \left( 
\left(\Box \phi\right)^2 \phi_{\mu} \phi^{\mu \nu} \phi_{\nu} 
- 2\, \Box \phi\, \phi_{\mu}  \phi^{\mu \nu} \phi_{\nu \rho}
\phi^{\rho} - \phi_{\mu \nu} \phi^{\mu \nu} \phi_{\rho} \phi^{\rho \s}
\phi_{\s} + 2 \phi_{\mu} \phi^{\mu \nu} \phi_{\nu \rho} \phi^{\rho \s} \phi_{\s}  \right) \Big], 
\end{eqnarray}
$\phi_\m = \de_\m \phi$ and $F_4$, $F_5$ are arbitrary functions of 
$\phi, X$. These actions reduce to Galileonic actions in an appropriate decoupling limit, when gravity is turned off (this has been shown in \cite{Gleyzes:2014qga} using the results of \cite{Koyama:2013paa}). On the other hand, when gravity is fully dynamical, beyond Horndeski is characterised by equations of motion of order higher than two. Using diffeomorphism (diff.) invariance and selecting the unitary gauge, one can show that constraint equations exist that are able to avoid the propagation of additional degrees of freedom \cite{Gleyzes:2014qga, Lin:2014jga}.
 
Hence, one would be tempted to freely add the aforementioned beyond Horndeski Lagrangians to the Horndeski action of the previous subsection with arbitrary coefficients, so to form a more general scalar tensor theory. However, as anticipated in \cite{Langlois:2015cwa}, and as we shall further develop later on, issues arise that suggest that an additional propagating mode could appear in some cases. Such problems become manifest when one does {\it not} select the unitary gauge in which the scalar field depends only on time \cite{Deffayet:2015qwa}. In the following sections, treating the general system without selecting any gauge, we will make some  progress in understanding the genuine dynamics induced by beyond Horndeski theories.
 
Before starting our analysis we introduce a suitable geometrical formulation for these Lagrangians that greatly simplifies the calculation. The basic point of this interpretation can be already found in \cite{Gleyzes:2014qga}, but again the choice of the unitary gauge hides the proper construction.
The approach is to introduce quantities according to the
constant scalar field hypersurface, $\phi=$ const.
This hypersurface is characterised by the following geometrical quantities
\begin{equation}
\frac{A^{\mu}}{\sqrt{-A^2}}, \qquad P^{\mu}_{\,\,\nu}, \qquad \Phi^{\mu}_{\,\,\nu},
\end{equation}
where $-A^{\mu}/\sqrt{-A^2}$ is the unitary normal vector,
\begin{equation}
P^{\alpha}_{\,\mu} = \delta^{\alpha}_{\,\mu} - \frac{1}{A^2}
A_{\mu} A^{\alpha} \label{proj}
\end{equation}
the projection tensor on the hypersurface, and 
\begin{equation}
\Phi_{\,\mu}^{\nu} = - P_{\,\mu}^{\alpha} P_{\,\beta}^{\nu}
\Nb_{\alpha} A^{\beta}
\label{phiextrinsic}
\end{equation}
is the extrinsic curvature of the constant $\phi$ hypersurface multiplied by $\sqrt{-A^2}$.
Notice that these objects have nothing to do with the usual ones associated with a space-time foliation.

The beyond Horndeski Lagrangians can be expressed in terms of the extrinsic curvature $\Phi^{\mu}_{\,\a}$ in the following way:
\begin{eqnarray}
{}^{BH}{\cal L}_4 &=& X F_4(\phi, \,X)\, \delta^{\alpha}_{[\mu} \delta^{\beta}_{\nu]} \,
\Phi^{\mu}_{\alpha} \Phi^{\nu}_{\beta}\,,  \\
{}^{BH}{\cal L}_5 &=& - X F_5(\phi,\,X)\, \delta^{\alpha}_{[\mu} \delta^{\beta}_{\nu}  \delta^{\gamma}_{\rho]}\,
\Phi^{\mu}_{\alpha} \Phi^{\nu}_{\beta} 
\Phi^{\rho}_{\gamma}\,.
\end{eqnarray}
Using the expression for the extrinsic curvature (\ref{phiextrinsic}), this can be written as 
\begin{eqnarray}
{}^{BH} {\cal L}_4 &=& X F_4 (\phi,\,X)\, {\cal M}^{\alpha \beta}_{\mu \nu} \, \Nb_{\alpha} A^{\mu} \Nb_{\beta} A^{\nu} ,
\label{BH4-M} \\
{}^{BH} {\cal L}_5 &=& X F_5 (\phi,\,X) \, {\cal M}^{\alpha \beta \gamma}_{\mu \nu \rho}\, \Nb_{\alpha} A^{\mu} \Nb_{\beta} A^{\nu} \Nb_{\gamma} A^{\rho} , \label{BH5-M}
\end{eqnarray}
where 
\begin{equation}
{\cal M}^{\alpha \beta}_{\mu \nu} = P^{\alpha}_{[\mu} P^{\beta}_{\nu]}\,, 
\qquad 
{\cal M}^{\alpha \beta \gamma}_{\mu \nu \rho} = P^{\alpha}_{[\mu} P^{\beta}_{\nu}P^{\gamma}_{\rho]}\,.
\end{equation}
The matrix ${\cal M}$ has a crucial property for the arguments that we are going to develop:
\begin{equation}
{\cal M}^{\alpha ..\gamma}_{\mu ... \rho}  
A_{\alpha} A^{\mu} = {\cal M}^{\alpha ..\gamma}_{\mu ... \rho}  
A_{\gamma} A^{\rho} =0\,.
\end{equation}
This property stems from the fact that ${\cal M}$  is built in terms of the projection tensor (\ref{proj}). As we will show in section \ref{sec-prim}, this leads to the existence of a primary constraint necessary to avoid 
the propagation of an extra mode in beyond Horndeski, despite the fact that the equations of motion have higher derivatives.

We stress that the unitary gauge used in the literature \cite{Gleyzes:2014qga, Lin:2014jga} is a very special choice of gauge where the constant time hypersurface and the constant scalar field hypersurface coincide. The analysis in this special gauge could lead to misleading results, as was explicitly pointed out first in~\cite{Deffayet:2015qwa} and then in \cite{Langlois:2015cwa, Langlois:2015skt}. For this reason, we will refrain from making this choice in what comes next.

\section{Kinetic terms}\label{sec-kin}

\no In order to identify the kinetic terms and carry out the analysis of constraints we need to separate space and time, performing a 3+1 decomposition.
We introduce the time vector flow $t^\mu\,=\,\partial/\partial t$ decomposed as 
\be
t^\mu\,=\,N\,n^\mu+N^\mu \,,
\ee
where $n^\mu$ is the unit normal vector to the $t=$ const. hypersurface, $N$ the lapse function and $N^\mu$ the shift vector orthogonal to the normal vector.   
The constant time hypersurface is then characterised by the following three quantities:
\begin{equation}
n^{\mu}\,, \qquad h^{\mu}_{\,\,\nu}\,, \qquad K^{\mu}_{\,\,\nu}\,,
\end{equation}
where   $h^{\mu}_{\,\,\nu} =\delta^{\mu}_{\nu} + n^{\mu} n_{\nu}$ is the projection tensor on the hypersurface  and $K^{\mu}_{\,\,\nu}$ the associated extrinsic curvature
\be\label{ext-def}
K_{\mu \nu}\,=\,\frac{1}{2 N}\left( \dot{h}_{\mu\nu}-D_{(\mu} N_{\nu)} \right) \,.
\ee
With ``dot'' we mean the Lie derivative respect to $t^\mu$\,, $D_{\mu}$ is the 3D covariant derivative on the constant time hypersurface and the parenthesis $(\dots)$ on the indices denote symmetrisation.  
Following \cite{Langlois:2015cwa}, we decompose $A_{\mu}$ into the normal and transverse components with respect to the aforementioned hypersurface:
\begin{equation}
A_{\mu} = -A_* n_{\mu} + \hat{A}_{\nu} h^{\nu}_{\,\mu}\,.      
\label{hatA}
\end{equation}
The expression for the covariant derivative of $A_{\mu}$ can be decomposed into various pieces depending on the derivatives of its components and of the metric:
\bea \label{dmuanuc}
\nabla_{\mu} A_{ \nu} = 
D_{\mu} \hat A_{\nu}-A_*\, K_{\mu\nu}+n_{(\mu} \left( K_{\nu) \rho } \hat{A}^\rho-D_{\nu)}A_*\right)
+\,n_\mu n_\nu \left(V_*- \hat{A}_\rho \,a^\rho \right) \,,
\eea
where $a^\mu=n^\nu \,\nabla_{\nu} \,n^\mu$ is the acceleration vector.
In equation (\ref{dmuanuc}), as well as for the whole (beyond) Horndeski Lagrangian, time derivatives appear only for the three dimensional metric $h_{\m\n}$ (inside the extrinsic curvature) and for the component $A_*$ (inside what we called $V_*$). $V_*$ plays for $A_*$ the same role that $K_{\m\n}$ plays for $h_{\m\n}$, i.e.
\be
V_* \equiv n^{\mu} \nabla_{\mu} A_* = \frac{1}{N} \left( \dot{A}_* -N^\mu D_\mu A_* \right)\,.
\ee
It is convenient therefore to work directly with the extrinsic curvature and $V_*$, instead of the real velocities~$\dot{h}_{\m\n}$ and $\dot{A}_*$, identifying those terms as the kinetic contributions to the action\footnote{To avoid confusion, with {\it kinetic terms} we indicate contributions to the Lagrangian that contain time derivatives; while for quartic (beyond) Horndeski the kinetic terms are at most bilinear 
in the time derivatives of the fields, for quintic Horndeski they are at most trilinear.}.
This allows us to treat the fields in a decomposed space-time while still remaining in a covariant form.
Remember however that $V_*$ contains the second time derivative of the scalar field, hence it represents a potentially dangerous contribution that could lead to the propagation of the \Ost mode.

A further advantage of this procedure is that the Lagrangian densities do not depend explicitly on the lapse and shift functions. This is because such quantities are implicitly included in $K_{\,\,\nu}^\mu$ and $V_* $.
This is a huge simplification that considerably reduces the number of fields involved in the calculation.
Performing a standard ADM canonical analysis would be very complicated, as was already shown in~\cite{Deffayet:2015qwa} for the case of quartic beyond Horndeski only.
On the other hand, (beyond) Horndeski Lagrangians are diffeomorphism invariant, so intuitively we do not expect any modification to the algebra of constraints associated with the lapse and shift, as they are the generators of such a symmetry.
However, this is not straightforward to show and a general proof is still missing. Steps forward in this direction have been made very recently in \cite{Langlois:2015skt}, where for a simple quartic Lagrangian, this was indeed shown to be the case.  
  
For the purposes of this paper, it is sufficient to retain in the Lagrangian only the highest order terms in the extrinsic curvature, so we obtain the following expressions (we adopt the notation used in \cite{Langlois:2015cwa}):
\begin{eqnarray}
{\cal L}^{\rm kin}_4 
&=&  2 {\cal B}^{\alpha}_{\mu}\, V_*\, K^{\mu}_{\alpha}
+{\cal K}_{\mu \nu}^{\alpha \beta} K^{\mu}_{\alpha} K^{\nu}_{\beta}\,, \label{lagkin4} \\
{\cal L}^{\rm kin}_5 
&=& 3 {\cal B}^{\alpha \beta }_{\mu \nu }\, V_* \, K_{\alpha}^{\mu} K_{\beta}^{\nu} 
+{\cal K}_{\mu \nu \rho}^{\alpha \beta \gamma} K^{\mu}_{\alpha} K^{\nu}_{\beta}
K^{\rho}_{\gamma}\,.
\label{lagkin}
\end{eqnarray}
In the following, for simplicity, we also assume that the functions $G_4, G_5, F_4$ and $F_5$ depend only on $X$, and not on $\phi$. 

For Horndeski, we obtain for the quantities $ {\cal B}$ and ${\cal K}$ the following expressions
\begin{eqnarray}
{}^H {\cal B}^{\alpha}_{\mu} &=& 0, \qquad 
{}^H {\cal B}^{\alpha \beta}_{\mu \nu}=0
\label{AB-H}\,,
\\ 
{}^H {\cal K}_{\mu \nu}^{\alpha \beta}
&=& - G_4 \, h^{\alpha}_{[\mu} h^{\beta}_{\nu]} + 2 G_{4 X} \left( A^2  h^{\alpha}_{[\mu} h^{\beta}_{\nu]} 
- \hat{A}^2 \hat{P}^{\alpha}_{[\mu} \hat{P}^{\beta}_{\nu]} \right)
\label{K-H4}\,,
\\
{}^H {\cal K}_{\mu \nu \rho}^{\alpha \beta \gamma}
&=& \frac13 G_{5 X} A_* \left( A^2 h^{\alpha}_{[\mu} h^{\beta}_{\nu} h^{\g}_{\rho]} - \hat{A}^2 \hat{P}^{\alpha}_{[\mu} \hat{P}^{\beta}_{\nu} \hat{P}^{\g}_{\rho]} \right)
\label{K-H5}\,,
\end{eqnarray}
where 
\begin{equation}
\hat{P}^{\alpha}_{\,\mu} = h^{\alpha}_{\,\mu} - \frac{1}{\hat{A}^2}
\hat{A}_{\mu} \hat{A}^{\alpha}
\end{equation}
is the three-dimensional projection tensor defined in terms of $\hat{A}_{\mu}$.
Note that there are no terms containing $V_*$, since the ${\cal B}$ quantities of equation \eqref{AB-H} vanish. This ensures that the equations of motion are second order and that the Horndeski Lagrangian only propagates at most three degrees of freedom. This is achieved by the special Horndeski tuning between the (non-minimal) coupling of gravity to the derivatives of the scalar field.  While the expressions associated with the quartic Horndeski have already been derived in \cite{Langlois:2015cwa}, this is the first time these expressions have been found for quintic Horndeski. 

%
%
%
The relevant kinetic terms for beyond Horndeski are
\begin{eqnarray}
{}^{BH} {\cal B}^{\alpha}_{\mu} &=& 
 F_4 A_* \hat{A}^2 \hat{P}^{\alpha}_{\mu}\,, \qquad 
 {}^{BH} {\cal B}^{\alpha \beta}_{\mu \nu}=
  - F_5 A_*^2 \hat{A}^2 \hat{P}^{\alpha}_{[\mu} \hat{P}^{\beta}_{\nu]} \,,
\label{B-BH4-3D}  
  \\
{}^{BH} {\cal K}_{\mu \nu}^{\alpha \beta}
&=& - F_4 \left[ A^4 h^{\alpha}_{[\mu} h^{\beta}_{\nu]} - A^2 \hat{A}^2 \hat{P}^{\alpha}_{[\mu} \hat{P}^{\beta}_{\nu]} +2\hat{A}^4 \hat{P}^{\alpha}_{\mu} \hat{P}^{\beta}_{\nu} - \hat{A}^4 \left( \hat{P}^{\alpha}_{\mu} h^\beta_\nu + \hat{P}^{\b}_{\nu} h^\a_\mu \right) \right] \,,
\label{K-BH4-3D}  
\\
{}^{BH} {\cal K}_{\mu \nu \rho}^{\alpha \beta \gamma}
&=& F_5 A_* \left[A^4 h^{\alpha}_{[\mu} h^{\beta}_{\nu} h^{\g}_{\rho]} - A^2\hat{A}^2 \hat{P}^{\alpha}_{[\mu} \hat{P}^{\beta}_{\nu} \hat{P}^{\g}_{\rho]} + \hat{A}^4 \left( \hat{P}^{\alpha}_{[\mu} \hat{P}^{\beta}_{\nu]} \left( \hat{P}^{\g}_{\rho} - h^\g_\rho \right) + \text{sym.} \right)\right] \,,
\label{K-BH5-3D} 
\end{eqnarray}
where ``$\text{sym.}$'' in (\ref{K-BH5-3D}) stands for the symmetric permutation of doublets of vertical indices (e.g. the last term in eq. (\ref{K-BH4-3D})).
Since the mixing term only appears in beyond Horndeski theory, we will mostly omit the superscript ``BH'' from ${\cal B}^{\alpha}_{\mu}$ and  ${\cal B}^{\alpha \beta}_{\mu \nu}$.

For beyond Horndeski the quantities in eq.~\eqref{B-BH4-3D} do not vanish, therefore there are potentially dangerous mixings between $V_*$ (containing the second derivative of the scalar field) and $K_{\,\mu}^\nu$ (containing the first derivative of the spatial metric). Such contributions to the action lead to higher order equations of motion. This suggests, but not necessarily implies, the presence of additional propagating degrees of freedom. We now study the existence of a primary constraint that could prevent the propagation of an additional (ghost) mode.

\section{Primary constraints in beyond Horndeski theories}\label{sec-prim}

\no The natural tool for counting the number of degrees of freedom is the (Dirac) canonical analysis of constraints.
A complete analysis, however, is beyond the scope of this work and here we concentrate on studying the existence of primary constraints. In particular we focus on the constraint that is able to remove the \Ost mode, and assume that the four first class constraints associated with the diff. invariance still remain. 
Of course, the existence of a primary constraint is not enough to remove a physical dof, nevertheless this would be the first necessary condition for it. Moreover, to our knowledge, there are no known Lorentz invariant theories that propagate half degrees of freedom: this would be the case if there are an odd number of second class constraints.
A complete analysis for the quartic Lagrangian has been recently performed in \cite{Langlois:2015skt}, confirming that a secondary constraint does indeed exist.

Primary constraints exist when, passing to the Hamiltonian formalism,
all the velocities {\it cannot} be expressed in terms of the fields and their conjugate momenta.
This translates to relations (constraints) between the fields and momenta that need to be added to the canonical Hamiltonian through Lagrangian multipliers. 

As explained in section \ref{sec-kin}, instead of working with the true velocities, we work with closely related quantities and therefore define the conjugate momenta accordingly:
\be
\pi_* \equiv \frac{1}{\sqrt{-g}} \frac{\delta {S}}{\delta V_*} \,, \qquad
\pi^{\a}_{\m} \equiv \frac{1}{\sqrt{-g}}
\frac{\delta {S}}{\delta K^{\mu}_{\alpha}} \,.
\ee
Notice that this definition also differs from the usual one due to the presence of the factor $1/\sqrt{-g}$\,; this helps to completely remove the lapse from the relations.

In Horndeski theory, the primary constraint needed to remove the \Ost ghost is automatically enforced by its construction, and it is $\pi_* \approx 0$\,.\footnote{The customary  notation ``$\approx$'' means weak equality, i.e. equality on the phase space determined by constraints.}

\subsection{Beyond Horndeski} 

\no Using the expression for beyond Horndeski theories given in eqs.~(\ref{BH4-M}) and (\ref{BH5-M}), together with  the expression of $\nabla_{\alpha} A^{\mu}$ given in eq.~(\ref{dmuanuc}), the conjugate momenta are obtained as 
\begin{eqnarray}
\pi_* &=& 2 X F_4 \, {\cal M}^{\alpha \beta}_{\mu \nu} \,n_{\alpha} n^{\mu}\, \nabla_{\beta}A^{\nu} 
+3 X F_5\, {\cal M}^{\alpha \beta \gamma}_{\mu \nu \rho}\, n_{\alpha} n^{\mu}\, \nabla_{\beta}A^{\nu}\, \nabla_{\gamma}A^{\rho} \,, \label{mom1full}
\\
\hat{A}^{\m} \hat{A}_{\a} \pi^{\a}_{\m} &=& 2 X F_4  {\cal M}^{\alpha \beta}_{\mu \nu} 
\left(- A_* \hat{A}_{\alpha} \hat{A}^{\mu} + \hat{A}^2 n_{(\alpha} \hat{A}^{\mu)}  \right)
\nabla_{\beta}A^{\nu}  \nonumber\\
&& + 3 X F_5  {\cal M}^{\alpha \beta \gamma}_{\mu \nu \rho} 
\left(- A_* \hat{A}_{\alpha} \hat{A}^{\mu} + \hat{A}^2 n_{(\alpha} \hat{A}^{\mu)}  \right)
\nabla_{\beta}A^{\nu} \, \nabla_{\gamma}A^{\rho} \,. \label{mom2full}
\end{eqnarray}
Using the properties of the matrix ${\cal M}$
\begin{equation}
 {\cal M}^{\alpha .. \beta}_{\mu.. \nu} n_{\alpha} A^{\mu}= {\cal M}^{\alpha .. \beta}_{\mu.. \nu} \hat{A}_{\alpha} A^{\mu}=0 \,,
\end{equation}
which stem from the fact that ${\cal M}$ is constructed from the projection tensor $P^{\alpha}_{\mu}$ and $A^{\mu} = - A_* n^{\mu} + \hat{A}^{\mu}$, we can derive the following identities 
\begin{equation}
	{\cal M}^{\alpha.. \beta}_{\mu.. \nu}\, \hat{A}_{\alpha} \hat{A}^{\mu}
	= A_*^2\, {\cal M}^{\alpha.. \beta}_{\mu.. \nu}\, n_{\alpha} n^{\mu}\,, \qquad 
	{\cal M}^{\alpha.. \beta}_{\mu.. \nu}\, n_{\alpha} \hat{A}^{\mu} 
	=A_* \,{\cal M}^{\alpha.. \beta}_{\mu.. \nu}\, n_{\alpha} n^{\mu}\,\,.
\end{equation}
Hence, we can find a primary constraint of the form
\begin{equation}
A_* \left(2 \hat{A}^2 - A_*^2\right) \pi_* - \hat{A}^{\mu} \hat{A}_{\alpha}\, 
\pi^{\alpha}_{\mu} \,\approx\, 0 \,. 
\label{primary-BH}
\end{equation} 
With the formulation of beyond Horndeski theories given in section \ref{secBHT}, it is therefore very easy to show the existence of the primary constraint necessary to remove the \Ost ghost.
It is important to notice that the constraint (\ref{primary-BH}) is a linear combination of the conjugate momenta $\pi_*$ and $\pi^{\alpha}_{\mu}$. In this case it is possible to remove $V_*$ by a suitable field redefinition and the system can be recast as a second order one. However there is no guarantee that the new system will be Lorentz invariant.

In order to better understand the properties of the conjugate momenta (\ref{mom1full}), (\ref{mom2full}) and for the sake of the next subsection, we now explain the above result focusing only on the highest order terms in the extrinsic curvature as given in eq. (\ref{lagkin4}) and (\ref{lagkin}).
The conjugate momenta simplify to
\begin{eqnarray}
\pi_* &=& 2 {\cal B}^{\alpha}_{\mu} K^{\mu}_{\alpha} 
+ 3  {\cal B}^{\alpha \beta}_{\mu \nu} K^{\mu}_{\alpha} K^{\nu}_{\beta}, 
\label{mom1}
\\
\pi^{\alpha}_{\mu} &=& \left(2 {\cal B}^{\alpha}_{\mu} + 6 {\cal B}^{\alpha \beta}_{\mu \nu} K^{\nu}_{\beta}\right) V_*
+2 {\cal K}^{\alpha \beta}_{\mu  \nu} K^{\nu}_{\beta} 
+ 3  {\cal K}^{\alpha \beta \gamma}_{\mu  \nu \rho} K_{\nu}^{\beta } K^{\rho}_{\gamma}.
\label{mom2}
 \end{eqnarray}
From eqs. (\ref{mom1}), (\ref{mom2}) it becomes easy to see that the key property of these momenta is that $V_*$ appears only in $\pi^{\alpha}_{\mu}$. This implies that to build a constraint we need to eliminate $V_*$ taking a suitable linear combination of components of $\pi^{\alpha}_{\mu}$. Using the properties of the three dimensional projection tensor, i.e.
\begin{equation}
\hat{P}^{\a}_{\mu} \hat{A}_{\a} \hat{A}^{\mu}=0 \,, \qquad
h^{\alpha}_{[\mu} h^{\beta}_{\nu]} 
\hat{A}_{\alpha} \hat{A}^{\mu}=\hat{A}^2 \hat{P}^{\beta}_{\nu}\,, \qquad
h^{\alpha}_{[\mu} h^{\beta}_{\nu} h^{\g}_{\rho]}\hat{A}_{\alpha} \hat{A}^{\mu}= \hat{A}^2 \hat{P}^{\beta}_{[\nu} \hat{P}^{\g}_{\rho]} \,,
\end{equation}
we can show the following relations for ${\cal B}$ and ${\cal K}$ in beyond Horndeski theories
\begin{eqnarray}
{\cal B}^{\alpha}_{\mu} \hat{A}_{\alpha} \hat{A}^{\mu} &=&
{\cal B}^{\alpha \beta}_{\mu \nu}\hat{A}_{\alpha} \hat{A}^{\mu} =0 \,, 
\label{relations-AB1}
\\
{}^{BH}{\cal K}^{\alpha \beta}_{\mu \nu} \hat{A}_{\alpha} \hat{A}^{\mu}
&=& A_* ( 2 \hat{A}^2 - A_*^2) {\cal B}^{\beta}_{\nu} \,,
\label{relations-AB2}
\\
{}^{BH}{\cal K}^{\alpha \beta \gamma}_{\mu \nu \rho} \hat{A}_{\alpha} \hat{A}^{\mu}
&=& A_* ( 2 \hat{A}^2 - A_*^2) {\cal B}^{\beta \gamma}_{\nu \rho}\,.
\label{relations-AB3}
\end{eqnarray}
These results imply that we can eliminate $V_*$ by contracting $\pi^{\alpha}_{\mu}$ with $\hat{A}_{\alpha} \hat{A}^{\mu}$
\begin{equation}
\hat{A}^{\mu} \hat{A}_{\alpha}\, \pi^{\alpha}_{\mu}
= A_* \left(2 \hat{A}^2 - A_*^2\right) \left(2 \; {\cal B}^{\alpha}_{\mu} K^{\mu}_{\alpha} 
+ 3 \;  {\cal B}^{\alpha \beta}_{\mu \nu} K^{\mu}_{\alpha} K^{\nu}_{\beta} \right).
\end{equation}
Then, it is straightforward to verify the primary constraint (\ref{primary-BH}).

To conclude, let us give the redefinition of the extrinsic curvature that eliminate the cross term between $V_*$ and the extrinsic curvature
\begin{equation}
K^{\mu}_{\alpha} = \bar{K}^{\mu}_{\alpha} - \frac{V_*}{A_* (2 \hat{A}^2 - A_*^2)} 
 \hat{A}_{\alpha} \hat{A}^{\mu} \,.
\label{diag4}
\end{equation}
We discuss in Appendix \ref{app1} the cases when this redefinition is in fact a disformal transformation of the extrinsic curvature.

\subsection{Beyond Horndeski $+$ Horndeski}
\no We now consider what happens to the primary constraint found in beyond Horndeski, when we combine this theory with the Horndeski one.
We will show that, when mixing actions of different orders, generically the primary constraint of the kind (\ref{primary-BH}) is lost.
In this regard, it is enough to consider only the highest order terms in the extrinsic curvature.

${\cal B}$ and ${\cal K}$ in Horndeski theories obey the following relations
\begin{eqnarray}
{}^{H}{\cal K}^{\alpha \beta}_{\mu \nu} \hat{A}_{\alpha} \hat{A}^{\mu}
&=& \frac{2 G_{4 X} A^2 - G_4 }{F_4 A_*}\, {\cal B}^{\beta}_{\nu} \,,
\label{relations-AB2-BH}
\\
{}^{H}{\cal K}^{\alpha \beta \gamma}_{\mu \nu \rho} \hat{A}_{\alpha} \hat{A}^{\mu}
&=& - \frac{G_{5X} A^2} {3 F_5 A_*}\, {\cal B}^{\beta \gamma}_{\nu \rho}\,.
\label{relations-AB3-BH}
\end{eqnarray} 
Using these relations, the conjugate momenta in the presence of Horndeski contributions become  
\begin{eqnarray}
\pi_* &=& 2 \; {\cal B}^{\alpha}_{\mu} K^{\mu}_{\alpha} 
+ 3 \;  {\cal B}^{\alpha \beta}_{\mu \nu} K^{\mu}_{\alpha} K^{\nu}_{\beta} \,, 
\label{piAmix1}
\\[2ex]
\hat{A}^{\mu} \hat{A}_{\alpha} \pi^{\alpha}_{\mu} &=&  
A_* \left( 2 \hat{A}^2 - A_*^2 \right) \left( 2  {\cal B}^{\alpha}_{\mu} K^{\mu}_{\alpha}  +
 3  {\cal B}^{\alpha \beta}_{\mu \nu} K^{\mu}_{\alpha} K^{\nu}_{\beta} \right)   \nonumber
\\[1ex]
&& +\, \frac{2 \left(2 G_{4X} A^2 -G_4 \right)}{F_4 A_*}\,
 {\cal B}^{\a}_{\m}  K^{\m}_{\a}  
- \frac{G_{5X} A^2} {F_5 A_*} \, {\cal B}^{\alpha \beta}_{\mu \nu} K^{\mu}_{\alpha} K^{\nu}_{\beta} \,.
\label{degmix1}
\end{eqnarray}

Let us first consider the case in which we do not mix different orders.
In these cases the primary constraint still persists.
\begin{itemize}

\item Quartic Beyond Horndeski $+$ quartic Horndeski
\begin{equation}
 \left[A_* \left(2 \hat{A}^2 - A_*^2 \right) + \frac{2 G_{4 X} A^2 -G_4 }{F_4 A_*}  \right] \pi_* - \hat{A}^{\mu} \hat{A}_{\alpha}\,
 \pi^{\alpha}_{\mu} \,\approx\, 0 \,. 
\end{equation}

\item Quintic Beyond Horndeski $+$ quintic Horndeski
\begin{equation}
 \left[A_* \left(2 \hat{A}^2 - A_*^2 \right) - \frac{G_{5X} A^2} {3 F_5 A_*}  \right] \pi_* - \hat{A}^{\mu} \hat{A}_{\alpha}\,
 \pi^{\alpha}_{\mu} \,\approx\, 0 \,. 
\end{equation}

\end{itemize}
In the above cases, the redefinition of the extrinsic curvature that removes the mixing term is given respectively by 
\bea
K^{\mu}_{\alpha} &=& \bar{K}^{\mu}_{\alpha} - \frac{ F_4 A_* V_*}{ F_4 A_*^2 \left(2 \hat{A}^2 - A_*^2 \right) + 2 G_{4X} A^2 -G_4 } 
\hat{A}_{\alpha} \hat{A}^{\mu}\,.
\label{HBH4} \\[2ex]
K^{\mu}_{\alpha} &=& \bar{K}^{\mu}_{\alpha} - \frac{ 3F_5 A_* V_*}{ 3F_5 A_*^2 \left(2 \hat{A}^2 - A_*^2 \right) - G_{5X} A^2 } 
\hat{A}_{\alpha} \hat{A}^{\mu}\,.
\label{HBH5}
\eea

So far, so good. Problems arise when we combine together different orders.
This was already realised in \cite{Langlois:2015cwa} for the case of quintic beyond Horndeski~$+$ quartic Horndeski.
Here we also provide the tools to analyse what happens joining together quartic beyond Horndeski~$+$ quintic Horndeski, a system that is more technically challenging to deal with.
In both of these cases the primary constraint is lost.
%
%
Indeed, the presence of Horndeski terms of different order, obstructs any {\it linear} combination of momenta (\ref{piAmix1}) and (\ref{degmix1}) that compensates and sets to zero the coefficients of both $ {\cal B}^{\a}_{\m}  K^{\m}_{\a}$  and $ {\cal B}^{\alpha \beta}_{\mu \nu} K^{\mu}_{\alpha} K^{\nu}_{\beta}$
\footnote{The only exception is for $G_4=X^{1/2}$. For this particular value the quartic Horndeski contribution vanishes.
(We will meet this special case also in the next section).}.

Of course the absence of the primary constraint which is present in isolated beyond Horndeski theories, does not necessarily mean that a different constraint could not arise.
Surely it cannot come from a linear combination of momenta\footnote{Remember that $\hat{A}^{\mu} \hat{A}_{\alpha} \pi^{\alpha}_{\mu}$ is the only linear combination of components of $\pi^{\alpha}_{\mu}$ that does not contain $V_*$.},
so in order to check its real absence, one should be able to prove the invertibility of the non-linear system of matrices equations (\ref{piAmix1})--(\ref{degmix1}), which is not a simple task.

To summarise, beyond Horndeski theories do have the primary constraint needed to avoid the \Ost instability, however this same constraint is spoiled when they are combined with contributions from different orders of  Horndeski.

\section{Disformal transformation}\label{sec-disf}
\no
In the previous section we found that, due to the linear nature of the primary constraint in beyond Horndeski theories,
a suitable redefinition of the extrinsic curvature can recast the theory into a manifestly second order system. 
In this section (see also Appendix \ref{app1}) we study the relation between this field redefinition and a disformal transformation of the metric.

The (generalised) disformal transformation \cite{Bekenstein:1992pj} is given by 
\begin{equation}
\bar{g}_{\mu \nu} = g_{\mu \nu} +\Gamma(\phi,\,X) A_{\mu} A_{\nu} \,. \label{disftransf}
\end{equation}
In \cite{Bettoni:2013diz, Zumalacarregui:2013pma, Bettoni:2015wta, Gleyzes:2014qga, Sakstein:2015jca} it  was shown that Horndeski theory is mapped into itself by a disformal transformation if $\Gamma$ is a function of the scalar field only and that the $X$ dependence of $\Gamma$ generates beyond Horndeski terms. In the following, we do not choose any gauge for our analysis; for simplicity, we only consider the case where $\Gamma$ depends strictly on $X$.  

We can show that by a disformal transformation, the Horndeski action is mapped to a combination of Horndeski and beyond Horndeski actions of the same order
\begin{eqnarray}
{}^{H}{\cal \bar{L}}_4[\bar{G}_4] &=&  {}^{H}{\cal L}_4[G_4]  +   {}^{BH}{\cal L}_4[D_4] \,,
\label{disH4}
\\
{}^{H}{\cal \bar{L}}_5[\bar{G}_5] &=&  {}^{H}{\cal L}_5[G_5]  +   {}^{BH}{\cal L}_5[D_5] \,;
\label{disH5}
\end{eqnarray}
where 
\begin{eqnarray}
G_4 &=& \bar{G}_4  (1 + X \Gamma)^{1/2}\,, 
\qquad  D_4 = 
\frac{\Gamma_{X}}{(1 + X \Gamma)^{1/2}} \left(  
\bar{G}_4 - 
\frac{2 X \bar{G}_{4\bar{X}}}{1 + X \Gamma} 
\right)\,, \label{G4D4} \\[2ex]
G_5 &=& \frac{\bar{G}_5}{(1 + X \Gamma)^{1/2}} + \int \frac{\bar{G}_5 \left(\Gamma + X \Gamma_X \right)}{2(1 + X \Gamma)^{3/2}}\, dX \,, \qquad D_5= \frac{X \bar{G}_{5\bar{X}} \Gamma_X}{3(1 + X \Gamma)^{5/2}} \,. \label{G5D5}
\end{eqnarray}
On the other hand, beyond Horndeski is mapped into itself only, with a different overall function
\begin{eqnarray}
{}^{BH}{\cal \bar{L}}_4[\bar{F}_4] &=&  {}^{BH}{\cal L}_4[F_4] \,, \\ 
{}^{BH}{\cal \bar{L}}_5[\bar{F}_5] &=&  {}^{BH}{\cal L}_5[F_5]  \,;
\end{eqnarray}
where 
\begin{equation}
F_4 = \frac{\bar{F}_4}{ (1 + X \Gamma)^{5/2}}\,, \qquad F_5 = \frac{\bar{F}_5}{ (1 + X \Gamma)^{7/2}}\,. \label{F4F5}
\end{equation}

From these results, we can easily see that the combination of Horndeski and beyond Horndeski of the same order can be mapped to Horndeski. Indeed
\begin{eqnarray}
{}^{H}{\bar{\cal L}}_4[\bar{G}_4]  +   {}^{BH}{\bar{\cal L}}_4 [\bar{F}_4]
&=& {}^{H}{\cal L}_4[G_4] \,, \label{disHBH4} \\
{}^{H}{\bar{\cal L}}_5[\bar{G}_5]  +   {}^{BH}{\bar{\cal L}}_5 [\bar{F}_5]
&=& {}^{H}{\cal L}_5[G_5] \,, \label{disHBH5}
\end{eqnarray}
for a disformal transformation $\Gamma(X)$ which satisfies 
\cite{Gleyzes:2014qga}
\begin{eqnarray} 
\Gamma_{4X} &=& \frac{F_4}{X^2 F_4 + 2 X G_{4X} - G_4} \,,\label{eq:quartic disformal rel.} 
\label{dis4} \\[2ex]
\Gamma_{5X} &=& \frac{3\,F_5}{3\,X^2 F_5 - X G_{5X}}\,, \label{dis5}
\end{eqnarray}
where $G_4,\, G_5$ are respectively given in equations (\ref{G4D4}), (\ref{G5D5}) and $F_4,\, F_5$ are given in (\ref{F4F5}).
Thus, the absence of the \Ost mode is ensured by the fact that these combinations can be mapped to Horndeski. In the absence of a coupling with matter, the theory is just Horndeski.
It is possible to check that, for (\ref{dis4})--(\ref{dis5}), the disformal transformation reproduces the redefinition of the extrinsic curvature that removes its mixing with $V_*$, i.e. eqs.~(\ref{HBH4}), (\ref{HBH5}).

On the other hand, if there is {\it only} beyond Horndeski, the disformal transformation cannot map the theory to Horndeski. In fact in this limit $\Gamma_{4,5}=-1/X$ and the disformal transformation becomes 
\begin{equation}
\bar{g}_{\mu \nu} = g_{\mu \nu} - \frac{1}{X} A_{\mu} A_{\nu} \,.
\end{equation}
This is  the projection tensor $P^{\nu}_{\mu}$ which satisfies $P^{\nu}_{\mu} A_{\nu}=0$, implying that it has a null eigenvalue and therefore is not invertible. Thus the disformal transformation required to map a combination of Horndeski and beyond Horndeski to Horndeski, given by eqs.~(\ref{dis4})--(\ref{dis5}), becomes singular in the limit $G_{4,5} \to 0$.\footnote{There is also another value of $G_4$ that makes the transformation singular, i.e. $G_4=X^{1/2}$. This clarifies that the spacial case encountered in the former section, is actually related to an ill-defined disformal transformation.}
This is consistent with the fact that beyond Horndeski is mapped into itself by the generalised disformal transformation. Surprisingly, as we showed in the previous section, there still exists a redefinition of the extrinsic curvature that removes the mixing (dangerous) term, eq.~(\ref{diag4}). However, it can be shown (see Appendix \ref{app1}) that this transformation is not a disformal transformation of the extrinsic curvature.
Accordingly to the generality of the disformal transformation \cite{Bekenstein:1992pj},
this implies that, when recast into a second order theory, beyond Horndeski is different from Horndeski.
It follows naturally therefore that beyond Horndeski is equivalent (up to a field redefinition) to a Lorentz breaking second order theory.

An important remark is that\footnote{Notice that $X$ transforms as $\bar{X} = X/(1+ X \Gamma)$.}, for the particular cases $F_4(X) = X^{-5/2}$ and $F_5(X) = X^{-7/2}$,
the generalised disformal transformation with an arbitrary function $\Gamma(X)$ is a {\it symmetry} of the theories of beyond Horndeski.
It would be interesting to investigate whether the existence of this symmetry is related to a first class constraint, i.e. if it is a gauge symmetry.


\section{Discussion}\label{sec-disc}
\no In this paper we examined the number of degrees of freedom present in the theory of beyond Horndeski, in a gauge-invariant way. 
We first rewrote beyond Horndeski in terms of the extrinsic curvature of the constant scalar field hypersurface;
then we derived the relevant kinetic terms for both beyond Horndeski and Horndeski theories.
Unlike Horndeski, in beyond Horndeski there is a mixing term between the second time derivative of the scalar field and the extrinsic curvature, which could give rise to the propagation of an additional (ghost) degree of freedom in absence of proper constraints.

We found that in beyond Horndeski there exists a primary constraint that is able to remove the \Ost instability.
Moreover this constraint occurs in the form of a linear combination of conjugate momenta.
This allowed us to identify the transformation of the extrinsic curvature that eliminates the aforementioned mixing term.

This primary constraint still persists when we join together theories of beyond Horndeski and Horndeski of the same order.
However, when mixing theories of different orders, the very same constraint is lost, as shown in \cite{Langlois:2015cwa} for the case of quintic beyond Horndeski $+$ quartic Horndeski.
Here we were able to show that this happens also in the other case, i.e. quartic beyond Horndeski $+$ quintic Horndeski.
Nevertheless, the loss of that constraint does not exclude the possibility that, in these cases, a different (non-linear) primary constraint could still arise. The study of this possibility however is well beyond the scope of this work.

Notice that in Ref.~\cite{Deffayet:2015qwa} a non-covariant method for removing the higher order time derivatives from the field equations was found. This is valid even for mixed order beyond Horndeski and Horndeski combinations which would seem to contradict our results.  However, the fact that a higher order theory can be recast in a second order one in a non-covariant way, is not necessarily connected with the number of propagating dof. Nothing forbids that one of the (second order) modes is indeed a ghost, like in the \Ost approach\footnote{We would like to thank C. Deffayet for discussions on this point.}.
The Hamiltonian analysis confirms itself as the unique tool for the correct counting of dof.

\begin{figure}[ht]
	\centering{
		\includegraphics[width=9cm]{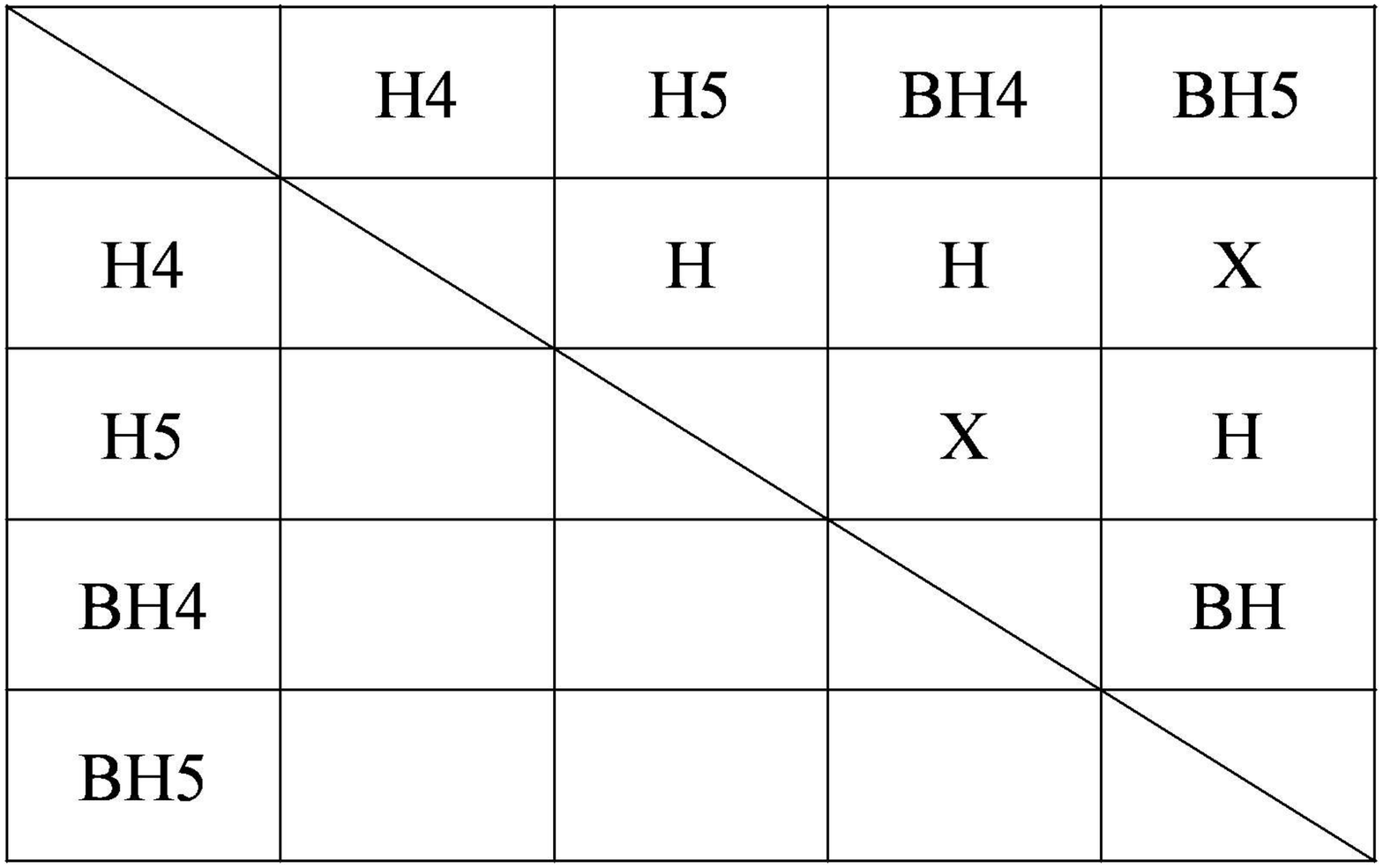}
	}
	\caption{Summary of the results on the combination of Horndeski and beyond Horndeski terms. H indicates that the theory can be mapped to Horndeski by a disformal tranformation while BH indicates that the disformal transformation maps it to beyond Horndeski. Crosses indicate theories that cannot be joined together without loosing the linear primary constraint.}
	\label{TAB}
\end{figure}


To identify the origin of the primary constraint, we studied the transformation of Horndeski and beyond Horndeski under a generalised disformal transformation which depends on the scalar field kinetic term, without fixing any gauge.
The results of this analysis are summarised in figure~\ref{TAB}.
We confirmed that Horndeski is disformally mapped to Horndeski plus beyond Horndeski of the same order \cite{Gleyzes:2014qga}.
On the other hand, beyond Horndeski is mapped to itself by the generalised disformal transformation.
Thus, in the absence of coupling with matter, the mixture of Horndeski and beyond Horndeski of the same order is nothing but Horndeski itself.
We checked that the redefinition of the extrinsic curvature that removes the mixing term can be derived from the disformal transformation only when there are both Horndeski and beyond Horndeski together. 
On the other hand, the disformal transformation that maps Horndeski plus beyond Horndeski to Horndeski is singular precisely in the limit that would leave the theory of beyond Horndeski alone. 
This is consistent with the fact that beyond Horndeski is mapped to beyond Horndeski itself.
In this case the transformation to remove the mixing term cannot be derived from the disformal transformation. 
This indicates that beyond Horndeski is disformally disconnected from Horndenski. 

The fact that beyond Horndeski cannot be mapped into Horndeski, makes this theory a covariant scalar-tensor theory that can be recast into a second order system which is not Horndeski, therefore it should be Lorentz breaking. 
It is also interesting that the primary constraint is preserved only when we include a Horndeski term of the same order, but it is spoiled when a different order Horndeski term is added.
In the first case however the theory is just Horndeski up to a generalised disformal transformation.

It will be important to continue the canonical analysis, performing the dynamical evolution of the primary constraint.
For the quartic case it has been recently shown that it leads to a secondary second class constraint \cite{Langlois:2015skt}; for the quintic case however such an analysis is still missing.
Moreover, it would also be interesting to study couplings with matter. We will come back to these issues in a forthcoming work. 

\section*{Acknowledgments}
We would like to thank Cedric Deffayet, David Langlois, Gustavo Niz and Jeremy Sakstein for useful discussions. MH is supported by a U.K. Science and Technology Facilities Council (STFC) research studentship. KK is supported by the UK Science and Technology Facilities Council grants ST/K00090/1 and the European Research Council grant through 646702 (CosTesGrav).

\appendix

\section{Redefinition of $K$ and disformal transformation}
\label{app1}

\no In this Appendix we examine the relation between the transformation of the extrinsic curvature that removes the mixing term between itself and $V_*$ (eq. (\ref{diag4})), and the transformation of the extrinsic curvature under a generalised disformal transformation of the metric (\ref{disftransf}). 
For simplicity, and to avoid very long formulae that are not illuminating, we only give the expression of the conformal piece and of the term proportional to $V_*$. It is indeed this last one that plays the crucial role.

The disformal transformation acts on the extrinsic curvature in the following way
\be
\bar{K}_{\mu\nu} = B\, K_{\mu\nu} - C\, V_* \left( \hat{A}_\mu + \hat{A}^2\Gamma A_\mu \right)\left( \hat{A}_\nu + \hat{A}^2\Gamma A_\nu \right) \,, \label{Ktransf}
\ee
where
\be
B = \left(\frac{1 + \hat{A}^2 \Gamma}{1 + A^2 \Gamma}\right)^{1/2} \,, \qquad C = \frac{B\, A_*\, \Gamma_X}{(1 + \hat{A}^2 \Gamma)^2} \,. \label{Cdisf}
\ee

In order to compare this transformation with the one coming from the analysis of the primary constraint (\ref{diag4}),
we first need to slightly manipulate the latter.
Indeed, the transformation given in section \ref{sec-prim} is not unique, we can always multiply the new extrinsic curvature by a generic function of $X$ without changing its properties.
Moreover, since the disformal transformation acts on all the fields, and not just on the extrinsic curvature, we also need to accordingly transform the other objects present in  (\ref{diag4}).
Once this has been done, we obtain that the transformation coming from the primary constraint is indeed of the form (\ref{Ktransf}), but with
\be
B = \left(\frac{1 + \hat{A}^2 \Gamma}{1 + A^2 \Gamma}\right)^{1/2} \,, \qquad C = \frac{B\, A_*\, \Gamma_X}{(1 + \hat{A}^2 \Gamma)^2} - \frac{B}{A_* \left( A_*^2 - 2 \hat{A}^2(1+A^2)\Gamma \right)} \,. \label{Cdiag}
\ee

Clearly in (\ref{Cdiag}) there is an additional piece proportional to $V_*$ that it is not present for the disformal transformation. This is the key term.
When we add together beyond Horndeski $+$ Horndeski of the same order, using (\ref{dis4})--(\ref{dis5}) in the 
analogous equation of (\ref{Cdiag}) coming from (\ref{HBH4})--(\ref{HBH5}), this extra piece vanishes and the transformation reduces to (\ref{Cdisf}).
On the other hand with beyond Horndeski alone this extra term survives making the two transformations different.
We can therefore conclude that, since the transformation that is able to recast beyond Horndeski into a second order theory is not a disformal transformation, beyond Horndeski is equivalent to a Lorentz breaking second order theory.
Notice that the singular limit encountered in section \ref{sec-disf}, i.e. $\Gamma_{4,5}=-1/X$, manifests also here: indeed for this value the extra piece does not cancel out as one would expect.

\end{document}